\documentclass[english,preprint,endfloats,floatfix,showpacs,amsmath,amssymb,aps]{revtex4}
\usepackage[T1]{fontenc}
\usepackage[latin1]{inputenc}
\usepackage{graphicx}
\usepackage{amssymb}

\makeatletter

\@ifundefined{textcolor}{}
{%
 \definecolor{BLACK}{gray}{0}
 \definecolor{WHITE}{gray}{1}
 \definecolor{RED}{rgb}{1,0,0}
 \definecolor{GREEN}{rgb}{0,1,0}
 \definecolor{BLUE}{rgb}{0,0,1}
 \definecolor{CYAN}{cmyk}{1,0,0,0}
 \definecolor{MAGENTA}{cmyk}{0,1,0,0}
 \definecolor{YELLOW}{cmyk}{0,0,1,0}
 }

\@ifundefined{definecolor}
 {\usepackage{color}}{}

\makeatletter

\@ifundefined{textcolor}{}
{%
 \definecolor{BLACK}{gray}{0}
 \definecolor{WHITE}{gray}{1}
 \definecolor{RED}{rgb}{1,0,0}
 \definecolor{GREEN}{rgb}{0,1,0}
 \definecolor{BLUE}{rgb}{0,0,1}
 \definecolor{CYAN}{cmyk}{1,0,0,0}
 \definecolor{MAGENTA}{cmyk}{0,1,0,0}
 \definecolor{YELLOW}{cmyk}{0,0,1,0}
 }

\makeatletter

\@ifundefined{definecolor}
 {\@ifundefined{definecolor}
 {\usepackage{color}}{}
}{} \@ifundefined{definecolor}
 {\@ifundefined{definecolor}
 {\usepackage{color}}{}
}{} \makeatother


\usepackage{babel}

\makeatother

\usepackage{babel}

\begin{document}
\widetext \draft

\title{Statistics and scaling properties of temperature field in symmetrical
non-Oberbeck-Boussinesq turbulent convection }

\author{Yuri Burnishev and Victor Steinberg }

\address{Department of Physics of Complex Systems, The Weizmann Institute
of Science, Rehovot 76100, Israel}

\date{\today}
\begin{abstract}
The influence of symmetrical non-Oberbeck-Boussinesq (SNOB) effect on statistical and scaling properties of temperature field  in turbulent convection is investigated experimentally in $SF_{6}$ in the vicinity of its gas-liquid critical point (CP). The main conclusion of the studies is that besides the strong $Ra$ and $Pr$ dependence of the rms of temperature fluctuations  normalized by the temperature difference across the cell, different from the Oberbeck-Boussinesq (OB) case of turbulent convection, all rest of statistical and scaling properties of temperature field discussed in details are the same as in the OB case.
\end{abstract}

\pacs{47.27.-i, 44.25.+f, 47.27.Te, 47.20.Bp}

\maketitle
%\begin{multicols}{2}
%\narrowtext

\section{Introduction}

Our recent publication \cite{yuri} introduces a new type of Rayleigh-Benard turbulent convection, namely symmetric non-Oberbeck-Boussinesq (SNOB) turbulent convection. The latter is characterized by strong temperature
and density dependencies of thermodynamic and kinetic properties of a
supercritical fluid near its gas-liquid critical point (CP) at the average critical density of the fluid in the cell. This strong temperature and density dependence results in a strong
but symmetric height dependence
of the main physical properties, which enter into the expressions for  the control
parameters of turbulent convection. So, in spite of strong variations of the fluid
properties across the cell height, up-down symmetry
of the temperature drops across the top and bottom halves of the cell and
of the top and bottom thermal boundary layer widths is preserved.
Thus, it was shown \cite{yuri} that in this case the
same scaling of $Nu=F(Ra,Pr)$ with the Rayleigh number, $Ra=\beta gL^{3}\Delta/\nu\kappa$,  as in the Oberbeck-Boussinesq (OB) turbulent convection is preserved but a much stronger the Prandtl number, $Pr=\nu/\kappa$, dependence of $Nu$ is found. Here $g$ is the gravity acceleration, $\beta$ is the fluid isobaric thermal
expansion coefficient, $\nu$ and $\kappa$ are the fluid kinematic
viscosity and thermal diffusivity, respectively, $Nu=QL/\lambda\Delta$ describes the heat transport by turbulent convection, $Q$ is the heat flux density, $L$ is the the cell height, $\lambda$ is the thermal conductivity of the fluid, $\Delta=T_b-T_t$ is the temperature difference across the cell, and $T_b$ and $T_t$ are the temperatures of the bottom and top plates, respectively. In order to single out the influence of the non-OB effect on the heat transport, the heat transport experiments we repeated for each $Pr$ with an eight-fold larger non-OB effect by decreasing the cell height at the same $Ra$. As the result, the $Nu=F(Ra,Pr)$ scaling was not altered. Therefore, the conclusion has been made that the strong SNOB effect by itself is not responsible for the strong $Pr$ dependence
of the heat transport near CP observed in the experiment but, probably, strongly enhanced compressibility, which accompanies an increase of $Pr$ due to approach to CP, could cause the observed $Pr$ dependence.

 Next natural question arises whether the SNOB effect results in the same scaling relations as in the OB case
 for statistics of temperature and velocity fields. In our early papers \cite{shay1,shay2}, we have already studied the statistics, frequency power spectra and scaling of velocity and temperature fields. However, this old experiment was less controlled and had lower resolution in temperature measurements with a narrower range of $Pr$ and $Ra$ than in the current experiment. The goal of the experiments presented in this paper is to find out whether the SNOB effect in turbulent convection modifies the scaling properties of the characteristic frequencies, corresponding to large scale circulation, to the Bolgiano length and to the dissipation scale, of the temperature power spectra and correlation functions, and  the scaling properties of the structure functions of the temperature increments compared with the OB case. These statistical properties of the temperature field are studied in a wide range of $Pr$ and $Ra$, the same as was reported in our recent paper on the heat transport \cite{yuri}.

The outline of the paper is as follows: In Section II the experimental setup is described in details. In Section III the local temperature measurements and statistics of temperature filed are presented. Frequency spectra of temperature fluctuations and auto- and cross-correlation functions of temperature field are discussed in Section IV, and temperature structure functions at high values of $Ra$ and $Pr$ are shown in Section V. The results are summarized and discussed in section VI.

\section{Experimental setup}

As convecting fluid, high purity (99.998\%) gaseous sulfurhexafluoride $SF_{6}$ was
used in the single-phase region in the vicinity of the gas-liquid
critical temperature $T_{c}$ and at the critical density $\rho_{c}$.
This fluid was chosen due to its
relatively low $T_{c}$ and critical pressure $P_{c}$, and its
well-known thermodynamic and kinetic properties, both far away and
in the close vicinity of the CP. The region of the phase diagram exploited in our
experiment is located at $\rho_{av}=\rho_{c}$ and in the region of
the reduced mid-plane temperature,
$({T_{m}}-T_{c})/T_{c}$, between $1.6\times10^{-4}$ and
$3.4\times10^{-2}$. As we pointed out in Ref. \cite{yuri}, NIST \cite{nist} gives the following values for the critical  parameters:
$T_{c}=318.733$~K, $P_{c}=37.5455$~bars,
$\rho_{c}=743.81$~kg/m$^{3}$, which we further use in our
analysis. In Ref. \cite{yuri} we provided detail information about thermodynamic and kinetic properties of $SF_6$ in the region of our interest and the way of temperature and pressure calibration and determination of the critical density $\rho_c$.

The experiments were carried out in two cells: one cell of a square
cross section $a\times a=70\times70$ mm$^{2}$ and height $L=90$ mm
(aspect ratio $\Gamma=L/a=1.286$), and one cylindrical cell with diameter $D=90$ mm
and height $L=90$ mm ($\Gamma=L/D=1$).
The side walls of all cells were made of 4~mm thick plexiglass
($\lambda_{pl}=0.2$~W/m~K). The cells were enclosed into a high-pressure
stainless steel vessel. The interior of the cell communicated with
the gap between cell and pressure vessel, through a small hole of
about 1 mm located in the cell mid-plane. For the cylindrical cell,
this gap was narrower than 5 mm and tightly filled with glass wool
($\lambda_{gw}=0.04$~W/m~K), tightened by a band. The cell was
then inserted into the vessel with a sufficient force. In the square
cross section cells, the outer space was larger, but also tightly
packed with glass wool. In both cases the filler was used to reduce
the amount of gas surrounding the cell, in order to prevent convection
outside the cell and to reduce the heat transfer due to the outer
gas (see Fig. 1).

In the rectangular cell, a top plate made of $h_{t}=20$~mm thick stainless steel 304
($\lambda_{ss}=25.8$~W/m~K) and bottom plate made of $h_{b}=8$~mm
thick aluminium 7075 ($\lambda_{Al}=133$~W/m~K) were used. In the
cylindrical cell, copper plates for both top and bottom were
used. Heat was supplied at the cell bottom by a metal-film heater of resistance
100~$\Omega$, that covered uniformly the entire active area of the
bottom plate. The experiment was carried out at a constant heat
flux, which was measured during the experiment.

The high-pressure vessel itself was immersed into a water bath. The
top plate was cooled down by water circulating in a two loop temperature
controlled system. The first water loop was refrigerated by a circulator
(Lauda Inc.) with a temperature stability of $\pm10$~mK. The second
water circuit, coupled via a heat exchanger to the first one, cooled
down the pressure vessel. The water temperature in the second loop
was controlled by a heater located just before the inlet to the apparatus,
computer-controlled by a feedback loop using a thermistor placed in
the mixing chamber of the water bath. The entire apparatus was covered
with insulating polyuretane sheets to reduce thermal losses. The temperature
stability of the bath achieved in a such way was better than $0.3$
mK rms. All heat transfer and temperature measurements were carried
out at fixed volume of the gas at the critical density. To perform the local temperature
measurements, first the average temperature
of the water bath, $T_{bath}$ was adjusted above $T_{c}$. The bottom
temperature in the cell, $T_{b}$ was then increased and $T_{bath}$
decreased in a such way that the temperature in the cell mid-plane
$T_{m}=(T_{b}+T_{t})/2$ remained constant. This way, the value of
$Pr$ at the cell midplane also remained constant with an accuracy
better than 1\% for a given temperature difference $\Delta=T_{b}-T_{t}$
across the cell or for a given $Ra$. Each time, the value of $T_{t}$
was recalculated from $T_{bath}$ and compared with the temperature
measured by the thermistor glued on the surface of the top plate.

Temperature measurements inside the cell were conducted using several thermistors
and one platinum thermometer of 1~$k\Omega$ resistance. The latter
provided the thermodynamic scale which defines the critical
parameters of the fluid, entering in either the parametric equation
of state \cite{sengers} or the tabulated data from NIST \cite{nist}.
Two glass-embedded stable thermistors P20BB204M from Thermometrics
Inc. were used in the water bath and inside the bottom plate. Three
glass bead thermistors B05KB204M of $150$~$\mu$m diameter from
Thermometrics Inc. were used to measure temperature inside the cell:
one was glued close to the center of the top plate and two were
mounted on the vertical motor-driven probe at the cell center.
  All thermistors together with the resistance thermometer were
calibrated in the thermally-regulated water bath (with a temperature
stability of $\pm10$~mK) against the secondary standard
platinum-resistance thermometer of 1~$k\Omega$ from the Russian
Institute of Standards. This thermometer was supplied with a
calibration accurate to $\pm10$~mK, on the ITS90 temperature scale.
Long-term stability (during 10 hours) of the thermistors was better
than 0.1 mK.

Pressure was measured with a calibrated pressure transducer TJE/727-23
(0-1000 psi) from Sensotec Inc. with a full scale output of
4.9975~V, which was read by a 6.5 digits multimeter. The pressure transducer
was calibrated against the absolute high precision pressure gauge
Heise Inc, USA (100.00 bar) with accuracy $\pm25$~mbar in the whole
range. Long-term stability (during 10 hours) of the pressure transducer
was better than 0.1 mbar.

The determination of the critical density of $SF_{6}$, which is
crucial for the experiment near CP on the critical isochore, was
based on the equation of state \cite{sengers} and on the NIST
data \cite{nist}. We adjusted the gas density far away from CP but
in the region still covered by the equation of state and the NIST
data \cite{sengers,nist}, by variation of T and P and their precise
measurements to be equal to $\rho_{c}$ according to either the
equation of state or the NIST data \cite{nist}. This method has an
error in $\rho_{c}$ due to the errors in T and P of the order of
$0.4\%$, while the accuracy in determination of $T_{c}$ and $P_{c}$
via the equation of state was completely defined by the errors in
the absolute values of $T$ and $P$.

\section{Local temperature measurements and statistics of temperature field in Rayleigh-Benard turbulent convection}

Local temperature measurements inside the cell were conducted by two
glass-bead thermistors B05KB204M of $150$~$\mu$m diameter and fast response time of 5 msec (immersed in water) from Thermometrics Inc., mounted on a stainless steel capillary tube
of 0.8 mm outer diameter, separated vertically by a fixed distance of about half of the
cell height (about 45 mm). Each thermistor is mounted on the copper
leads of 0.1 mm diameter and 5 mm long and suspended on its
contacting leads of 0.02 mm diameter and about 5 mm long, made of
platinum alloy ($\lambda=71.6$ W/m K) (Fig. 1). The
probe was driven vertically by a computer-controlled stepping motor at the cell center in both cylindrical
 and rectangular cells, allowing precise computer control of the thermistor positions. Besides in the rectangular cell another probe was used at one of the cell corners.
Temperature of the thermistors on the probe was measured in two ways. One way was the resistance measurements via an ac bridge and Lock-in amplifier (PAR model 124) at 83 Hz and A-to-D card. The frequency was chosen to be about twice larger than the expected highest frequency in the temperature frequency spectra at the highest values of $Ra\approx 10^{16}$, that is about 30-40 Hz. Another way was the thermistor resistance measurements via a 6.5 digits multimeter Keithley 2010 with a sampling rate of 54 Hz. The comparison showed a good agreement between two data sets measured by two techniques. At each value of $Ra$ and $Pr$ the temperature measurements were taken at up to 30 locations but the data only at three heights are presented in details: 10, 22.5 and 45 (center) mm with large statistics in average up to $10^7$ data points for each set.
 In the cylindrical cell, the data for 6 different $Pr=11,25,74,130,180,244$ and  up to 5 values of $Ra$ for each $Pr$ value were taken. So all together 66 data sets for different values of $Pr$, $Ra$, and 3 locations (10, 22.5 and 45 mm) for each set of $Pr,Ra$ were analyzed. In the rectangular cell, the data only for 3 values of $Pr=4.8,12,20$ and  5 values of $Ra$ for each $Pr$  and for the same three locations as above were analyzed.
Figure 2 presents segments of temperature fluctuations time traces $\delta T(t)=T(t)-\bar{T}$ for different $Ra$ and $Pr$ at a mid-height and at the center of the cylindrical and rectangular cells. Here $\bar{T}$ is the time average temperature. It can be seen that positive and negative spikes are distributed approximately equally. It is naturally reflected in almost symmetrical probability distribution functions (PDF) of temperature fluctuations at the central location in the both geometry cells with exponential tails taken at the same values of $Ra$ and $Pr$ as in Fig. 2 (Fig. 3). To compare PDFs at different heights at the cell center, we present PDFs of temperature fluctuations at several values of $Pr$ and selected values of $Ra$ in the both cylindrical and rectangular cells in Fig. 4. As one expects, out of the mid-height PDFs show significant skewness and exponential tails, as a rule, in the direction of positive temperature fluctuations. The rms temperature fluctuations normalized by the temperature difference across the cell, $T_{rms}/\Delta $, and  measured at the cell mid-height as a function of $Ra$ at different values of $Pr$ are shown in Fig. 5. The data are fitted by the power law $T_{rms}/\Delta =(46\pm 11) Ra^{\alpha}Pr^{\beta}$, where $\alpha=-0.43\pm 0.03$ and $\beta=1.24\pm 0.01$. The $Ra$ dependence is much stronger than $\alpha\approx -0.145$ found in the OB case \cite{castaing,libchaber,tong} while the $Pr$ dependence was never reported. The third $M_3$ and fourth $M_4$ moments of the temperature distributions at the cell mid-height as  a function of $Ra$ for different $Pr$ are presented in Fig. 6. Significant scatter between zero and unity is observed at $M_3$ and between 3 and 4 for $M_4$ without visible dependence on either $Ra$ or $Pr$ that witnesses on systematic deviations from the Gaussian and almost symmetrical distribution. Figure 7a,b,c shows a dependence of three moments of PDFs of temperature fluctuations on height $h$ at $Pr=130$ and three values of $Ra$: $4.6\times 10^{13},1.9\times 10^{14},6.1\times 10^{14}$. The normalized rms $T_{rms}/\Delta$ is independent of $h$ at $Pr=130$ and three values of $Ra$  as presented in Fig. 7a. The same conclusion can be made for all values of $Ra$ and $Pr$ explored in the experiment. The skewness (third moment) $M_3$ and flatness (fourth moment) $M_4$ show positive deviations towards 1 and 4, respectively, at the top and bottom of the cell, as one can see in Fig. 7b,c for $Pr=130$ and three values of $Ra$. The same conclusion can be reached for all values of $Ra$ and $Pr$ explored in the experiment.

\section{ Temperature power spectra, auto- and cross-correlation functions of the temperature field}

Further by using frequency power spectra and auto- and cross-correlation functions of temperature fluctuations we are going to study scaling of the frequency of the large scale temperature oscillations with $Ra$ and $Pr$. These frequency peaks in the power spectra and the corresponding characteristic oscillation periods in the correlation functions were reported in many studies on OB turbulent convection \cite{castaing,sano,cioni,tong1}. We concentrate first on the temperature measurements at the center and  the mid-height of the cell in a wide range of $Ra$ and $Pr$. As an example, we present in Fig. 8 the frequency power spectra of temperature $P(f)$ at $Pr=126$ and 3 values of $Ra=3.1\times 10^{13}, 1.1\times 10^{14}, 4\times 10^{14}$ and at $h=45$ mm in the cylindrical cell. A striking feature of all frequency power spectra studied is an emergence of either one (in a rectangular cell one observes either one main peak or the main peak and a much smaller second harmonic peak) or two sharp peaks at low frequencies. The presence of the second harmonic peak at $2f_c$, besides the main peak at $f_c$, indicates a nonlinear nature of the oscillations. The second peak can be comparable and sometime even larger (in cylindrical cell) than the main one (see Fig. 8). In our early measurements, the main peak in the frequency power spectra for temperature fluctuations was barely observed, probably, due to lower resolution and smaller statistics, whereas similar peaks in the frequency power spectra of velocity fluctuations were detected and analyzed \cite{shay1}. The normalized main peak frequencies for both cells at all values of $Pr$ and $Ra$ at $h=45$ mm are scaled as $f_{c}L^2/\nu=(0.53\pm 0.05)Ra^{0.43\pm 0.01}Pr^{-0.74\pm 0.03}$ (see Fig. 9) with the coefficient found from the high resolution compensating plot, which is presented in the inset in Fig. 9 (we used the value $\nu=5\times 10^{-4}$ cm$^2$/s, see Ref. \cite{yuri}). Good agreement with the result on scaling of the frequency of the oscillations found from the peak in the velocity power spectra reported in Ref. \cite{shay1} is found. Similar scaling relations were obtained for the main peak  $f_{c}L^2/\nu$ in the frequency power spectra at the cell center and other two heights below the cell mid-plane $h=22.5$ and $h=10$ mm in the same range of $Pr$ and $Ra$. As shown in Fig. 10 a,b, the corresponding scaling relations are $f_{c}L^2/\nu=(0.3\pm 0.05)Ra^{0.45\pm 0.02 }Pr^{-0.74\pm 0.07}$ and $f_{c}L^2/\nu=(1\pm 0.15) Ra^{0.41\pm 0.02}Pr^{-0.74\pm 0.07}$ at $h=22.5$ and $h=10$ mm, respectively, with close values of exponents for all three heights. Figure 11 shows an independence of $f_cL^2/\nu$ on the height at $Pr=11,Ra=3\times 10^{11}$ and $Pr=74, Ra=1\times 10^{14}$, as examples.

Similar information can be found from the temperature auto-correlation function (ACF) $g_{a}(\tau)=\lim_{\Upsilon\rightarrow\infty}\int_{0}^{\Upsilon}{\delta T(t)\delta T(t+\tau)dt}/\Delta^2$ as a function of the delay time $\tau$. Here $\delta T(t)=T(t)-\bar{T}$ is the temperature fluctuation and $\bar{T}$ is the average local temperature. Figure 12 presents several examples of the measured ACF at $h=45$ mm and at the following values of ($Pr,Ra$): ($25,2.1\times 10^{12}$), ($130,5.8\times 10^{14}$), and ($244,1.1\times 10^{15}$). First, the measured ACF exhibit decaying oscillations. Second, the characteristic oscillation period $\tau_c$ strongly depends on $Pr$ and $Ra$. And third, the characteristic oscillation period $\tau_c$ is equal to the inverse frequency $f_c$ of the main peak in the corresponding power spectrum, and the exponents in the scaling relations of $\tau_c^{-1}L^2/\nu$  and $f_cL^2/\nu$ with $Ra$ and $Pr$ are the same inside the error bars.

Similar features are also observed in the temperature cross-correlation function (CCF) $g_{c}(\tau)=\lim_{\Upsilon\rightarrow\infty}\int_{0}^{\Upsilon}{\delta T_1(t)\delta T_2(t+\tau)dt}/\Delta^2$ as a function of $\tau$. In the measurements, the two thermistors separated by a fixed distance of 45 mm apart are located on the same vertical temperature probe, which is driven vertically at the cell center. Figure 13 presents CCF at one location $h=22.5$ mm of the lower thermistor (and correspondingly of the upper one at $h=67.5$ mm ) and at three values of $Pr$ and $Ra$: ($11,4\times 10^{12}$), ($74,3\times 10^{13}$), ($180,2\times 10^{15}$). Similar to the auto-correlation functions, CCFs exhibit oscillations with the frequency $\tau_c^{-1}$ equal to the main peak frequency $f_c$ in the corresponding power spectrum. The oscillations have a decay with the coherence time $\tau_{dec}$, which is defined from the fit and is about an order of magnitude larger than the corresponding oscillation period $\tau_c$. Scaling relation of the normalized oscillation frequency $\tau_c^{-1}$ as a function of $Pr$ and $Ra$, found from the cross-correlation functions at $h=22.5$ mm is presented in Fig. 14. The power-law fit gives $\tau_{c}^{-1}L^2/\nu=(1\pm 0.08)Ra^{0.41\pm 0.02}Pr^{-0.72\pm 0.02}$ with the exponent values in a good agreement with those found for the main frequency peak in the power spectra (see Fig. 10a). We were not able to get the scaling relation for the coherence frequency $\tau_{dec}^{-1}L^2/\nu$ due to large scatter and insufficient number of the data points. On the other hand, one finds that the ratio $\tau_{dec}/\tau_c$ grows from approximately 3.5 at $Pr=11$ and $Ra=(1\div 4)\times 10^{12}$ till 4.5 at $Pr=74$ and $Ra=(3\times 10^{13}\div 1\times 10^{14})$ and up to 8.8 at $Pr=180$ and $Ra=(4\times 10^{14}\div 2\times 10^{15})$. It means that the growing number of the observed oscillation periods during the decay period indicates the growing coherence of the oscillation of the large scale circulation. As shown in Ref. \cite{tong1}, the large scale circulation occurs before the threshold for the coherent oscillations. There is a threshold for the onset of the coherent oscillations, which  are resulted from coherent action of the rising and falling plumes \cite{tong1}. Thus, our measurements show that the threshold value of $Ra$ for the onset of the coherent oscillations also depends on $Pr$.

Another feature of the cross-correlation function is the time delay (shift) $\tau_{sh}$ of its peak value from zero. Examples of the time delays are shown in Fig. 15 a,b for $Pr=244$ and three values of $Ra$ and for $Pr=130$ and another three values of $Ra$. This time delay is probably related to the large scale circulation, and so can supply information about its velocity. The $Ra$ and $Pr$ dependence of the normalized time delay is presented in Fig. 16, where the corresponding scaling relation is $\tau_{sh}^{-1} L^2/\nu=(12.1\pm 3.6)Ra^{0.34\pm 0.02}Pr^{-0.71\pm 0.01}$ with the coefficient determined from the high resolution compensating plot (see inset in Fig. 16). Relatively large scatter of the data, compared for example with the data in Figs. 9, 10, and 14, which is reflected in large error in the constant, could be probably the reason for the smaller value of the exponent in $Ra$ dependence 0.34, compared with that obtained for the normalized oscillation frequency $\tau_c L^2/\nu$ of the cross-correlation functions 0.41 (see Fig. 14 and the corresponding scaling relation above). As shown in Ref. \cite{tong1}, the large scale circulation velocity and the oscillation frequency are related to each other by simple relation, which in frequency domain gives the ratio 4. Due to different scaling exponents for $Ra$ dependence mentioned above, the ratio between $\tau_{sh}^{-1}$ and $\tau_c^{-1}$ varies between 1.3 and 2.2, which difference from 4 can also be caused by the large scatter and insufficient statistics.

To further examine the scaling properties of the frequency power spectra of temperature fluctuations we use the method introduced for hydrodynamic turbulence and suggested for convective turbulence in Ref. \cite{xia1}. The method utilizes the peak frequency $f_p$ corresponding to the maximum value of the temperature dissipation spectra $f^2P(f)$ as the characteristic frequency to collapse the power spectra for different $Ra$ and $Pr$. Figure 17 shows scaled frequency power spectra of temperature fluctuations $P(f)/P(f_p)$ versus $f/f_p$ for all values of $Pr$ and $Ra$ explored in the experiment in the cylindrical cell. An example of the dissipation spectrum $f^2P(f)$ at $Pr=244$ and $Ra=1.6\times 10^{13}$ at the height $h=45$ mm  and the cell center is presented in Fig. 18. The solid line in Fig. 18 is the sixth-order polynomial fit to the data, which allows an accurate determination of the frequency $f_p$ corresponding to the data maximum. It is clear from Fig. 17, that for each $Pr$ and $Ra$ there is well-defined $f_p$, with respect to which the frequency power spectrum is a universal function. Thus, we checked that the suggested method \cite{xia1} works well in a wide range of $Pr$ and $Ra$ up to their very high values. To compare the scaling results for $f_c$ and $f_p$, we present in Fig. 19 the normalized frequency $f_p L^2/\nu$ as a function of $Pr$ and $Ra$. The corresponding scaling relation is  $f_{p}L^2/\nu=(162\pm 21)Ra^{0.27\pm 0.01}Pr^{-0.43\pm 0.02}$ with the coefficient obtained from the compensating plot shown in the inset in Fig. 19. Thus we found experimentally  two characteristic frequencies, $f_c$ and $f_p$ in statistics and dynamics of the temperature field and studied their scaling relations in regards to the two control parameters of the problem $Pr$ and $Ra$.

Finally, the scaled frequency power spectra of the temperature fluctuations exhibit rather clear power-law region in the low frequency end with the exponent $1.4$ equal to the Bolgiano-Obukhov (BO) scaling though in the wave number domain that is more convincingly demonstrated by the high resolution compensating plot of $P(f)/P(f_p)(f/f_p)^{1.4}$ versus $f/f_p$ in the inset in Fig. 17. This is the only  clearly identified scaling region found in the scaled power spectra. On the other hand, it is rather well established \cite{cioni1,xia1}, that in developed turbulent convection at large values of $Ra$ and sufficiently small $Pr$, two scaling regions can be observed simultaneously. One of them is the BO region, where buoyancy controls the dynamics, and another is the inertial, Kolmogorov range, in which the effects of buoyancy are irrelevant \cite{monin}. These regions are separated in scale by the Bolgiano length scale $l_B$, and at scales $l>l_B$ the BO scaling is observed, while at $l<l_B$ the Kolmogorov scaling is found \cite{monin,cioni1,xia1}. According to \cite{chilla},  $l_B$ can be estimated by

\begin{equation}
l_B=L Nu^{1/2}(RaPr)^{-1/4},\label{1}
\end{equation}

and can be considered as a characteristic value of $l_B$.  Using the expression of Eq. (1), we can estimate the characteristic Bolgiano frequency based on large scale circulation velocity defined via $f_c$. Then the Bolgiano frequency can be written in the normalized form as

\begin{equation}
f_B(L^2/\nu)=(L^2/\nu)f_c L/l_B=0.64Ra^{0.53}Pr^{-0.39},\label{2}
\end{equation}

where $Nu=0.18 Ra^{0.3}Pr^{-0.2}$, taken from Ref. \cite{yuri}, is substituted into the expression of Eq. (1), and the expression for $f_c$ is used from the fit in Fig. 9. We point out that in the case considered the scaling of $f_B$ in both $Pr$ and $Ra$ is very different from the above found scaling for $f_p$ in contrast to Ref. \cite{xia1}. The ratio $f_B/f_p=3.9\times 10^{-3} Ra^{0.24}$ change rather strongly with $Ra$ but practically independent of $Pr$ in the range of $Pr$ and $Ra$ considered in the experiment. Finally, the largest characteristic frequency in the temperature power spectra is the dissipation one defined from the dissipation (or Kolmogorov) length \cite{monin,chilla}

\begin{equation}
l_d=L(PrRaNu)^{-1/4}=1.54Ra^{-0.325}Pr^{-0.2},\label{3}
\end{equation}

with $Nu$ substituted from Ref. \cite{yuri}. The corresponding dissipation frequency is $f_d(\nu/L^2)=0.42Ra^{0.75}Pr^{-0.54}$. In Fig. 17 we show by arrows for comparison the minimal and maximal values of $f_c/f_p$ and $f_B/f_p$ for some specified values of $Pr$ and $Ra$ in the whole range of these parameters studied, while $f_d/f_p$ remains outside of frequency range presented in the plot at all values of $Pr$ and $Ra$. From our estimates one finds that $f_B$  is located rather close to the intersect of $P(f)$ with the noise floor and therefore rather close to the cut-off frequency of the frequency power spectra for the most of the values of $Pr$ and $Ra$ considered in the experiment. Only for the lowest values of $Pr$ and $Ra$ one can expect to find $f_B$ close to $f_p$ (see Fig. 17).
To conclude, the experimental results for $f_c$ and $f_p$ and the theoretical estimates for $f_B$ and $f_d$ in the range of $Pr$ and $Ra$ considered in the experiment give the following range of changes for these characteristic frequencies: $f_c=[1.1\times 10^{-2}\div 0.13]$ Hz, $f_p=[0.3\div 1.5]$ Hz, $f_B=[1\div 75]$ Hz, and $f_d=[45\div 3\times 10^4]$ Hz. So, $f_d$ is out of scale of the power spectra obtained in the entire range of $Pr$ and $Ra$, where the cut-off frequency is exclusively determined by the noise floor, and $f_B$ for low $Pr$ and low $Ra$ probably could be found at frequencies significantly above the intersection with the noise floor.

To proceed further with the analysis of an experimental determination of $f_B$ and probable existence of the mentioned above two scaling regions we use a different approach based on temperature structure functions.

\section{Temperature structure functions in the Bolgiano-Obukhov regime of thermal convection at high $Pr$ and $Ra$.}

According to the values of $f_c$ and $f_p$ obtained experimentally and the values of $f_B$ obtained from estimates, one can expect at some values of $Pr$ and $Ra$ from the range of the values explored in the experiment to find $f_B$ rather close to $f_p$, between $f_c$ and the cut-off frequency, and therefore to observe the second scaling region at $f>f_B$ corresponding to the Kolmogorov scaling. As was shown in the recent publications \cite{benzi,benzi1}, a generalized extended self-similarity (ESS) method based on structure functions of temperature increments can be used to analyze the temperature data. Let us first to examine the behavior of the structure functions $S_p(\tau)=\langle|T(t+\tau)-T(t)|^{p}\rangle$ up to order $p=8$. Similar to the correlation functions discussed in the previous Section, the structure functions at least up to $p=4$ exhibit oscillations (Fig. 20), which minima are associated with one, two, etc. coherent oscillation periods of the large scale circulation. Naturally, the oscillation period and the scaling with $Pr$ and $Ra$ are the same as for the auto- and cross-correlation functions and the inverse frequency $f_c$ found from the power spectra (see Fig. 21).

To proceed further with the ESS method, let us use the approach, suggested in Ref. \cite{ching} and employed for the temperature data analysis in turbulent convection in Refs. \cite{xia1,sreeni}, and plot the ratio $S_1(\tau)/[S_2(\tau)]^{1/2}$ versus time increments $\tau$, as it is presented in Figs. 22 and 23 for several values of $Pr$ and $Ra$. On the plots in Fig. 22 a,b the data at (a) $Pr=11$ and $Ra=4\times 10^{11}$ and (b) $Pr=74$ and $Ra=1\times 10^{14}$ are shown with scaling in the entire BO region (indicated by the solid line) between $\tau_c$ and $\tau_B$ (found from the estimates) and with oscillations on scales $\tau\geq\tau_c$. The corresponding scaling exponent is close to 0.4 on the plot $S_2(\tau)$ (not shown) that corresponds to the exponent 1.4 found from the temperature power spectra (see Fig. 17). On the plots in Fig. 23 a,b for the data at (a) $Pr=180$ and $Ra=2\times 10^{15}$ and (b) $Pr=244$ and $Ra=3\times 10^{15}$ a change in the slope occurs at $\tau_B=0.59$ s and $\tau_B=0.93$ s, respectively, the time scales close to these found from the estimates. Two scaling ranges are identified above and below the breaking point, which we associate with $\tau_B$, corresponding to the Bolgiano length $l_B$. Thus, at $\tau>\tau_B$ the BO scaling regime is identified, while at $\tau<\tau_B$ the Kolmogorov scaling regime is found.

Figure 24 shows the structure functions $S_p$ up to order $p=8$ in the ESS presentation $S_p$ versus $S_2$ for $Pr=130$ and $Ra=2\times 10^{14}$ in the BO range of scales. Reasonable scaling is observed though it deteriorates slightly for the higher order structure functions. The normalized scaling exponents $\zeta_p/\zeta_2$ obtained from the plots in Fig. 24 are shown  in Fig. 25 in comparison with the results of averaging over all the data for all $Pr$ and $Ra$ studied in the experiment. We also show the data from Ref. \cite{sreeni} taken on turbulent convection in helium, the data on passive temperature fluctuations in turbulent flows in air \cite{ciliberto} and in helium \cite{tabeling} and theoretical suggestion \cite{leveque}. All results exhibit strong intermittent behavior of the temperature field with our data found surprisingly close to the theoretical curve corresponding to a passive scalar analysis.

\section{Discussion and Conclusions}

The detail studies of the statistical and scaling properties of the temperature fluctuation field of SNOB turbulent convection in a  wide range of $Pr$ and $Ra$ show that they are rather close to those investigated in the OB case. The only difference we found is in much stronger $Ra$ dependence of the normalized rms temperature fluctuations  $T_{rms}/\Delta$ than in the OB case and unusually strong $Pr$ dependence, which cannot be compared with the OB case due to lack of the data. On the other hand, scaling of the peak frequency of the coherent oscillations in the temperature power spectra as well as the corresponding periods of auto- and cross-correlation functions and structure functions agree well with the those in the OB case. Another visible difference found is the emergence of strong second harmonic peak sometimes even higher than the main one that points out on the strong nonlinearity of the coherent oscillations. We also found that the degree of coherence of the oscillations depends on $Pr$ besides the $Ra$ dependence studied in the OB case in Ref. \cite {tong1}. We verified that the method suggested in Ref. \cite{xia1} to collapse the temperature power spectra using the the peak frequency $f_p$ of the corresponding dissipation spectra works well in a wide range of $Pr$ and $Ra$ in SNOB turbulent convection though the corresponding scaling of $f_p$ differs significantly from the estimated scaling of $f_B$ in contrast to the results of Ref. \cite{xia1}. From the structure function analysis we establish that for the most of the range of $Pr$ and $Ra$ studied in the experiment only the BO scaling region is observed though for the smallest small values of $Pr$ and $Ra$ both scaling regions were identified. And finally, the dependence of the normalized scaling exponents of the structure functions $\zeta_p/\zeta_2$ on $p$ agrees rather well with those for the OB case and is found surprisingly close to the theoretical predictions for a passive scalar behavior \cite{leveque}.

This work is partially supported by grant from the Israel Science Foundation and by the Minerva Center for Nonlinear Physics of Complex Systems.

\begin{figure}
\includegraphics[width=9cm]{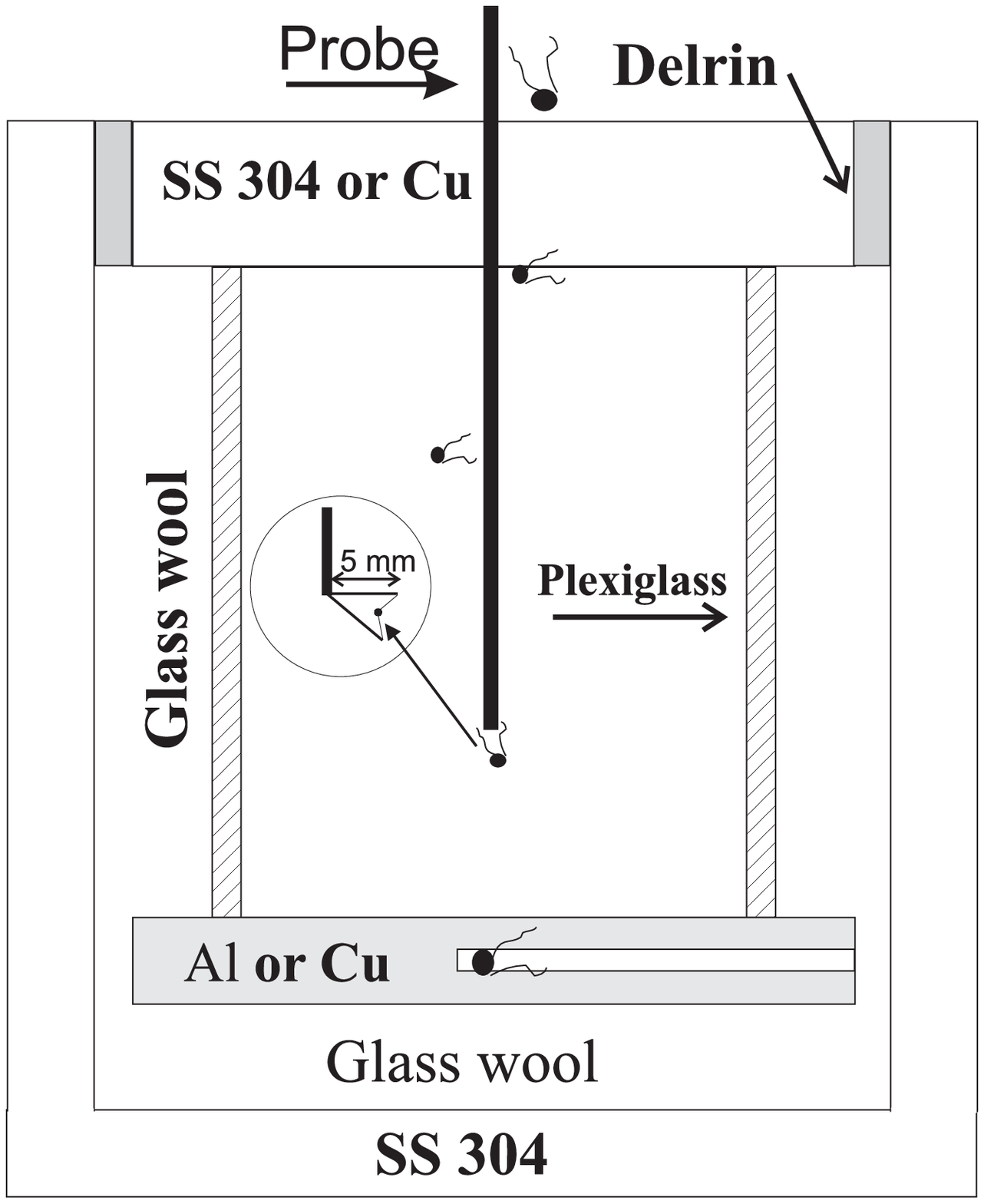}

\caption{Schematic drawings of a convective cell and of a thermistor probe for local temperature measurements.}

\label{fig:1}

\end{figure}
\begin{figure}
\includegraphics[width=9cm]{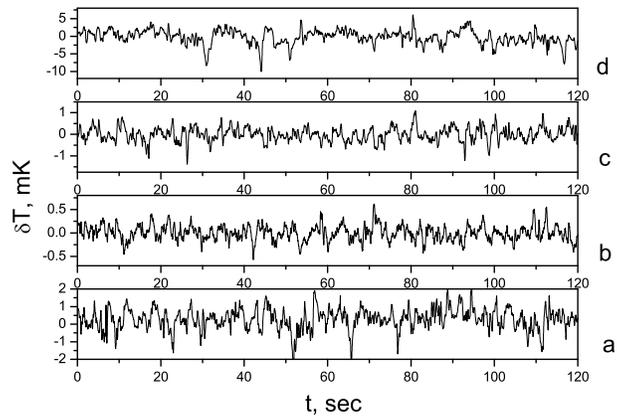}

\caption{Time series of temperature fluctuations $\delta T(t)=T(t)-\bar{T}$ at the cell center and at mid-height $h=45$ mm for the following values of $Pr$ and $Ra$: (a) (11, $4\times 10^{12}$); (b) (74, $1\times 10^{14}$); (c) (180, $2\times 10^{15}$); (d) (244, $3\times 10^{15}$).}

\label{fig:2}

\end{figure}

\begin{figure}
\includegraphics[width=9cm]{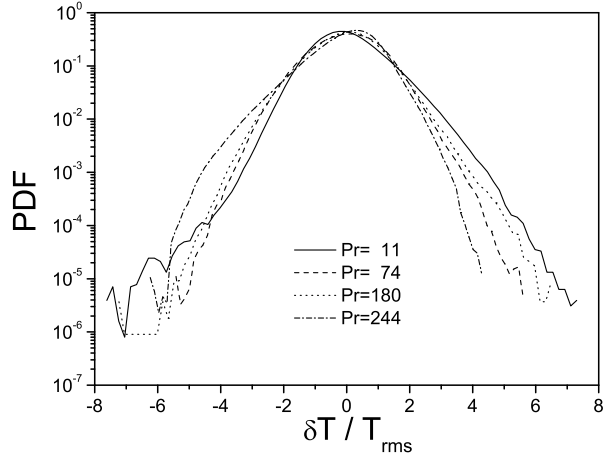}

\caption{PDFs of the normalized by the corresponding rms of temperature fluctuations $\delta T/T_{rms}$  at the cell center and at mid-height $h=45$ mm for the same values of $Pr$ and $Ra$ as in Fig. 2.}

\label{fig:3}

\end{figure}

\begin{figure}
\includegraphics[width=9cm]{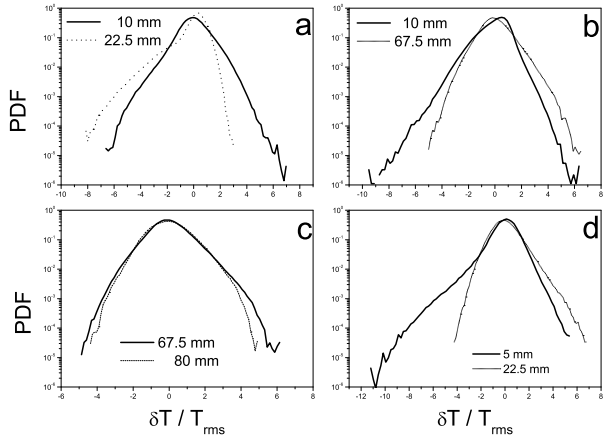}

\caption{PDFs of the normalized by the corresponding rms of temperature fluctuations $\delta T/T_{rms}$  at the cell center and at following values of heights $h$, $Pr$ and $Ra$: (a) (10 and 22.5 mm, 11, $4.7\times 10^{12}$); (b) (10 and 67.5 mm, 74, $1.1\times 10^{14}$); (c) (67.5 and 80 mm, 180, $2\times 10^{15}$); (d) (5 and 22.5 mm, 244, $3.7\times 10^{15}$).}

\label{fig:4}

\end{figure}

\begin{figure}
\includegraphics[width=9cm]{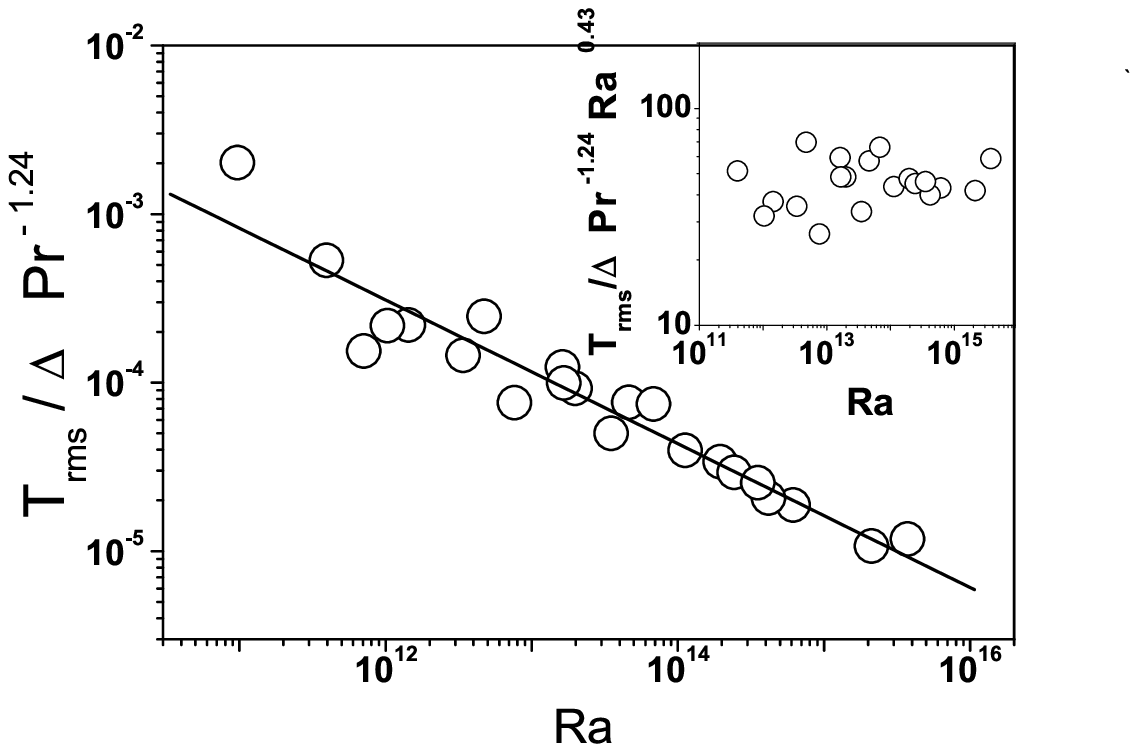}

\caption{Normalized rms of temperature fluctuations $T_{rms}/\Delta$ as a  function of $Ra$ for different $Pr$. The solid line is the fit to the data $T_{rms}/\Delta=(46\pm 11) Ra^{-0.43\pm 0.03}Pr^{1.24\pm 0.01}$. The inset shows the compensated plot of $(T_{rms}/\Delta)Ra^{0.43}Pr^{-1.24}$ versus $Ra$, from which the constant is found.}

\label{fig:5}

\end{figure}

\begin{figure}
\includegraphics[width=9cm]{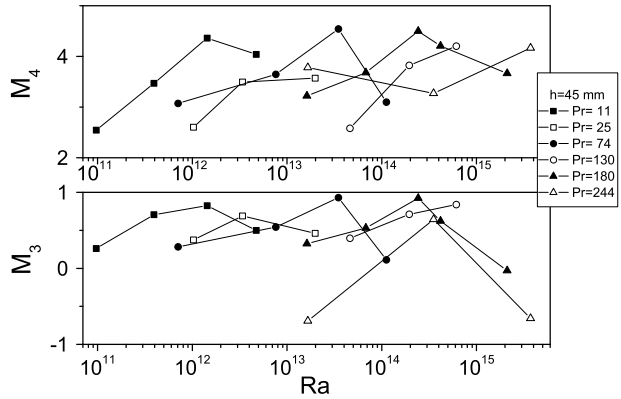}

\caption{Third (skewness) $M_3$  and fourth (flatness) $M_4$ moments of PDFs of temperature fluctuations in a whole range of $Pr$ and $Ra$ studied.}

\label{fig:6}

\end{figure}

\begin{figure}
\includegraphics[width=9cm]{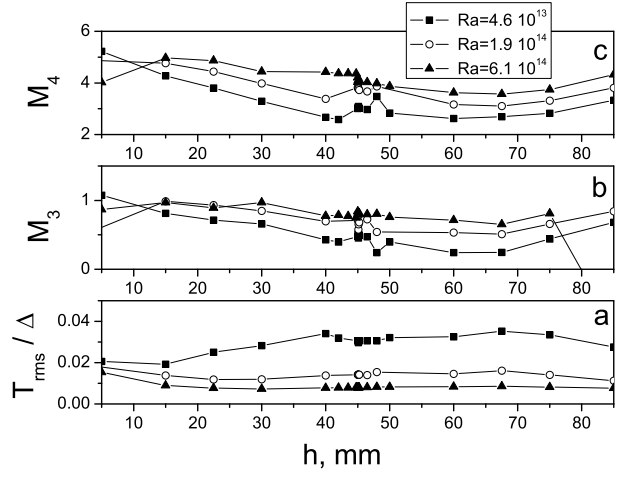}

\caption{Dependence of (a) the normalized dispersion $T_{rms}/\Delta$, (b) skewness $M_3$, and (c) flatness $M_4$ of PDFs of temperature fluctuations on $h$ at $Pr=130$ and three values of $Ra$: $4.6\times 10^{13}, 1.9\times 10^{14}, 6.1\times 10^{14}$. }

\label{fig:7}

\end{figure}

\begin{figure}
\includegraphics[width=9cm]{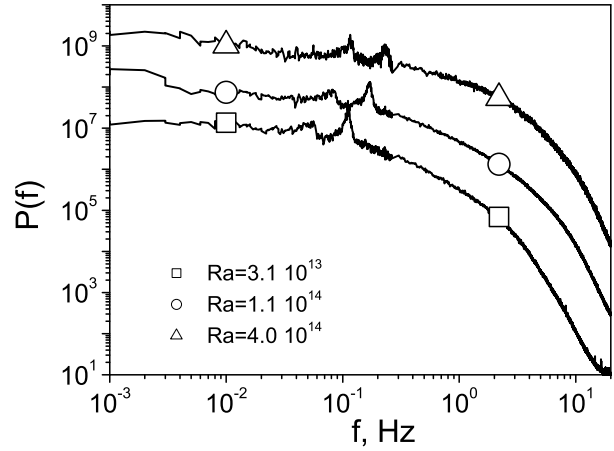}

\caption{Frequency power spectra of temperature fluctuations $P(f)$ at the cylindrical cell center and mid-height $h=45$ mm for $Pr=130$ and three values of $Ra$: $3.1\times 10^{13}, 1.1\times 10^{14}, 4\times 10^{14}$, where the second peak $2f$ in the power spectra sometime larger than the main one. }

\label{fig:8}

\end{figure}

\begin{figure}
\includegraphics[width=9cm]{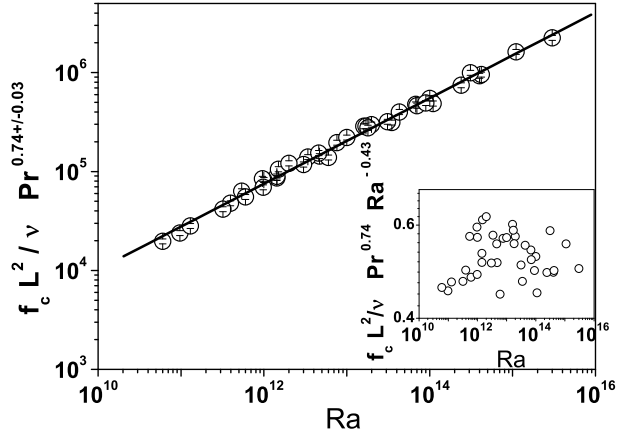}

\caption{Normalized frequency of the main peak in the power spectra at all values of $Pr$ and $Ra$ under studies at $h=45$ mm in both types of cells. The solid line is the fit for all the data $f_c L^2/\nu=(0.53\pm 0.05) Ra^{0.43\pm 0.01}Pr^{-0.74\pm 0.03}$.}

\label{fig:9}

\end{figure}

\begin{figure}
\includegraphics[width=9cm]{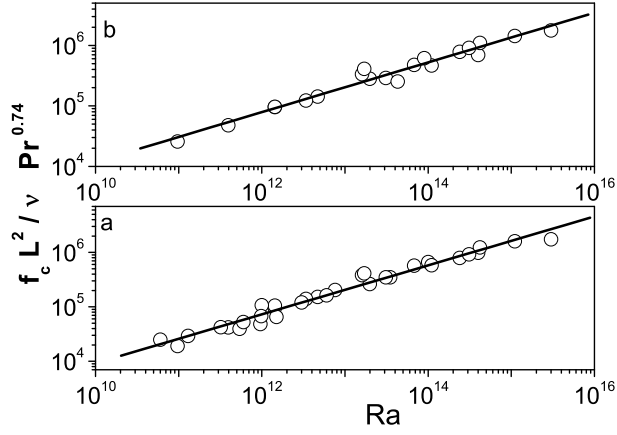}

\caption{Normalized frequency of the main peak in the power spectra $f_c L^2/\nu$ at all values of $Pr$ and $Ra$ under studies at (a) $h=22.5$ mm  and (b) $h=10$ mm  in both types of cells. The solid line are the fits for all the data (a) $f_c L^2/\nu=(0.3\pm 0.06) Ra^{0.45\pm 0.02}Pr^{-0.74\pm 0.07}$ and (b) $f_c L^2/\nu=(1\pm 0.15) Ra^{0.41\pm 0.02}Pr^{-0.74\pm 0.07}$.}

\label{fig:10}

\end{figure}

\begin{figure}
\includegraphics[width=9cm]{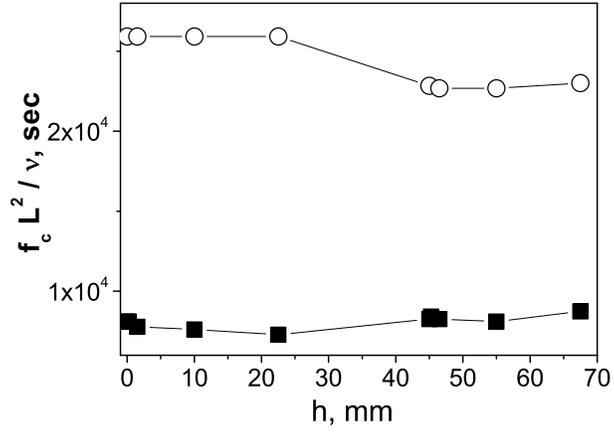}

\caption{Normalized frequency of the main peak in the power spectra $f_cL^2/\nu$ as a function of height $h$ at $Pr=11, Ra=3\times 10^{11}$ (lower data), $Pr=74, Ra=1\times 10^{14}$ (upper data).}

\label{fig:11}

\end{figure}

\begin{figure}
\includegraphics[width=9cm]{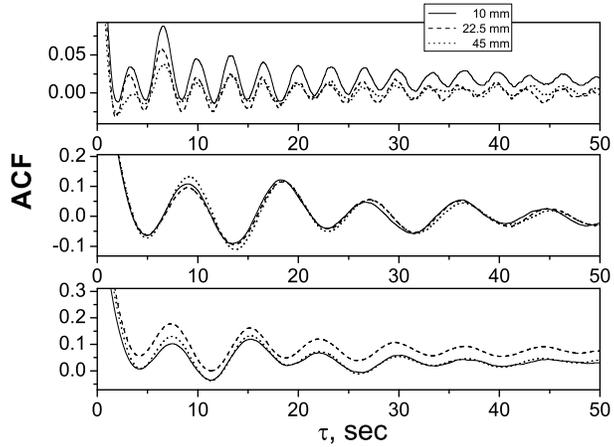}

\caption{Several examples of auto-correlation functions of temperature fluctuations measured at three locations $h=45, 22.5, 10$ mm and the following values of $Pr$ and $Ra$: (a) 25, $2.1\times 10^{12}$; (b) 130, $5.8\times 10^{14}$; (c) 244, $1.1\times 10^{15}$.}

\label{fig:12}

\end{figure}

\begin{figure}
\includegraphics[width=9cm]{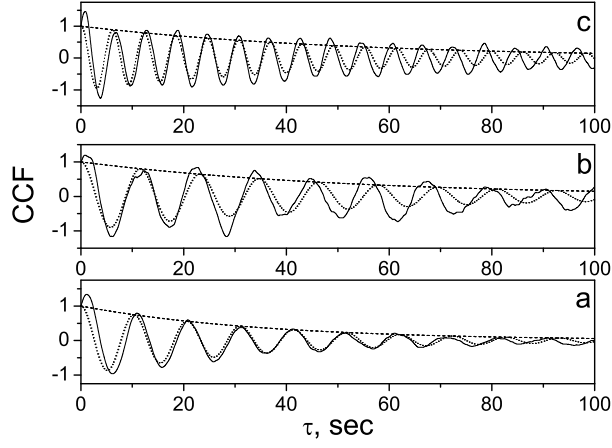}

\caption{Several examples of cross-correlation functions of temperature fluctuations measured between two locations at $h=22.5$, $h=67.5$ mm and the following values of $Pr$ and $Ra$: (a) 11, $4\times 10^{12}$; (b) 74, $3\times 10^{13}$; (c) 180, $2\times 10^{15}$. The dotted curve is the full fit and the dashed curve is the exponential fit to the decay.}

\label{fig:13}

\end{figure}

\begin{figure}
\includegraphics[width=9cm]{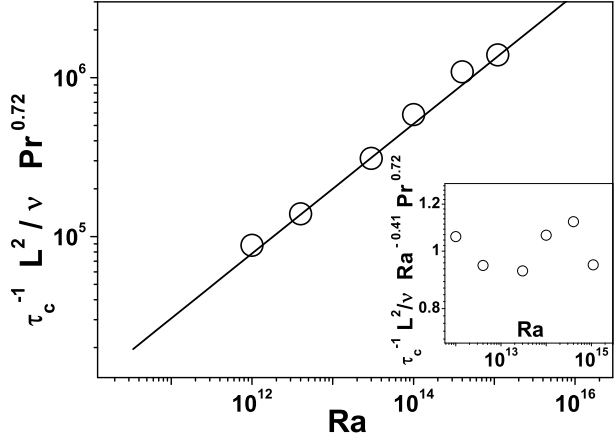}

\caption{Normalized frequency $\tau_c^{-1} L^2/\nu$ obtained from the cross-correlation functions as a function of $Pr$ and $Ra$ at $h=22.5$ mm. The data based on the cross-correlation functions are compared with those obtained from the frequency power spectra at $h=22.5$ mm and good agreement is found. The solid line is the fit $\tau_c^{-1}L^2/\nu=(1\pm 0.08) Ra^{0.41\pm 0.02}Pr^{-0.72\pm 0.02}$. }

\label{fig:14}

\end{figure}

\begin{figure}
\includegraphics[width=9cm]{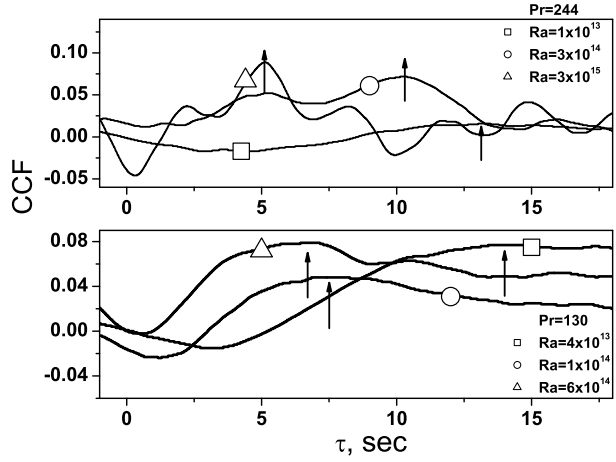}

\caption{Examples of the time delays of the peak of the cross-correlation functions at $Pr=244$ and three values of $Ra$ (upper plot) and at $Pr=130$ and another three values of $Ra$ (lower plot). }

\label{fig:15}

\end{figure}

\begin{figure}
\includegraphics[width=9cm]{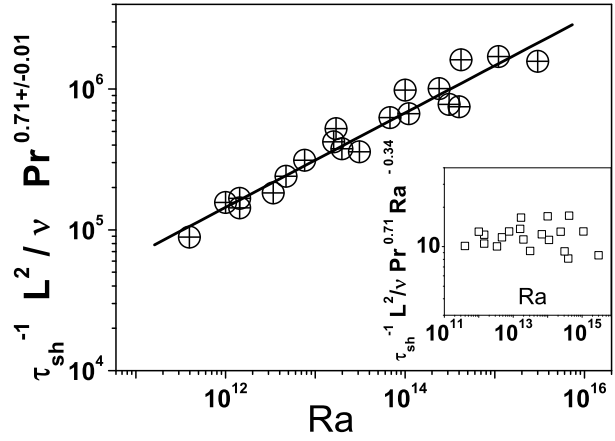}

\caption{Normalized frequency $\tau_{sh}^{-1} L^2/\nu$ obtained from the cross-correlation functions as a function of $Pr$ and $Ra$ at $h=22.5$ mm.  The solid line is the fit $\tau_{sh}^{-1}L^2/\nu=(12.1\pm 3.6) Ra^{0.34\pm 0.02}Pr^{-0.71\pm 0.01}$ with coefficient taken from the compensating plot in the inset. }

\label{fig:16}

\end{figure}

\begin{figure}
\includegraphics[width=9cm]{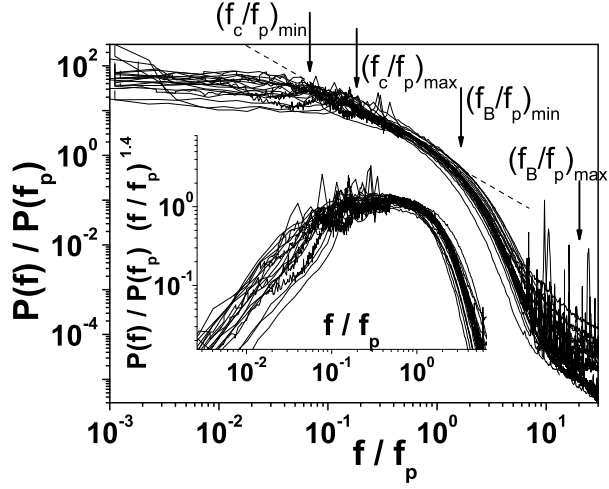}

\caption{ Scaled temperature power spectra taken  in the cylindrical cell at $h=45$ mm for all values of $Pr$ and $Ra$ explored in the experiment. Arrows indicate minimum and maximum values of $f_c/f_p$ and $f_B/f_p$: minimum value of $f_c/f_p\approx 0.07$ at $Pr=74$ and $Ra=7\times 10^{11}$; maximum value of $f_c/f_p\approx 0.18$ at $Pr=180$ and $Ra=2\times 10^{15}$; minimum value of $f_B/f_p\approx 1.66$ at $Pr=11$ and $Ra=9\times 10^{10}$; maximum value of $f_B/f_p\approx 20.2$ at $Pr=244$ and $Ra=3\times 10^{15}$. The dash line has a slope of $1.4$. The inset: the compensating plot $P(f)/P(f_p)(f/f_p)^{1.4}$ versus $f/f_p$.}

\label{fig:17}

\end{figure}

\begin{figure}
\includegraphics[width=9cm]{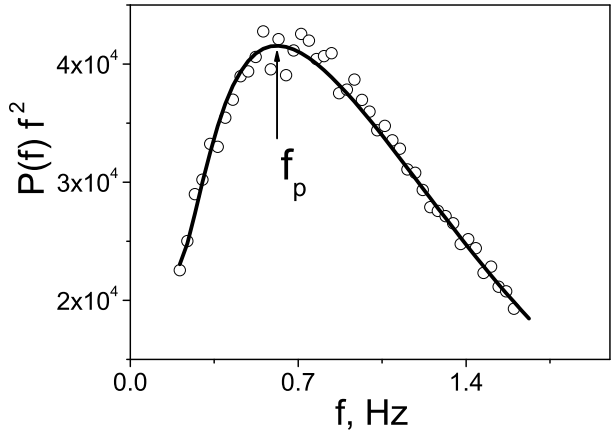}

\caption{An example of temperature dissipation spectrum $P(f)f^2$ versus $f$ taken in the cylindrical cell at $h=45$ mm at $Pr=244$ and $Ra=1.6\times 10^{13}$ is shown. The solid line is the sixth-order polynomial fit to determine the peak frequency $f_p$.}

\label{fig:18}

\end{figure}

\begin{figure}
\includegraphics[width=9cm]{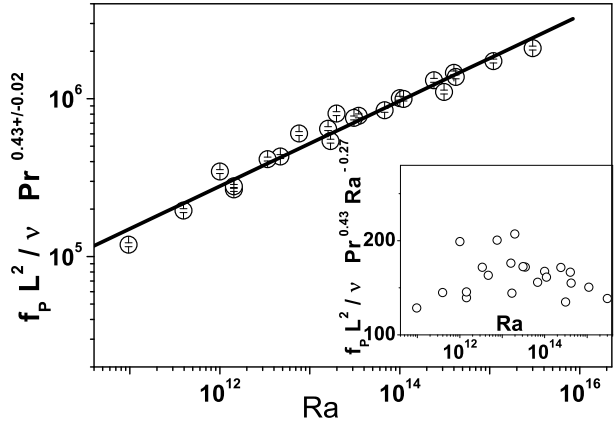}

\caption{Normalized frequency of the peak of the dissipation spectra as a function of $Ra$ at various $Pr$. The solid line is the fit $f_p L^2/\nu=(162\pm 21)Ra^{0.27\pm 0.01}Pr^{-0.43\pm 0.02}$.}

\label{fig:19}

\end{figure}

\begin{figure}
\includegraphics[width=9cm]{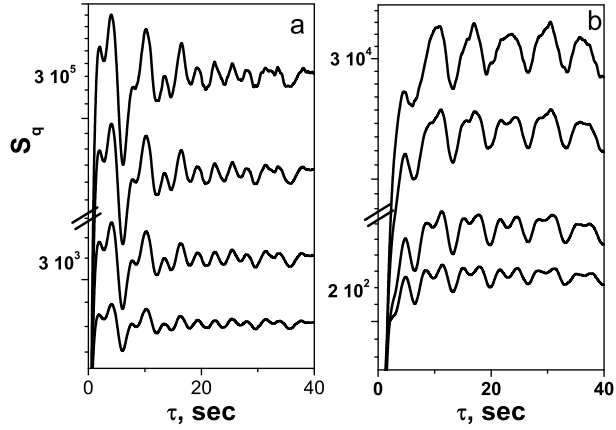}

\caption{Temperature structure functions $S_p$ up to $p=4$ for (a) $Pr=74$ and $Ra=1\times 10^{14}$; (b) $Pr=244$ and $Ra=3\times 10^{15}$.}

\label{fig:20}

\end{figure}

\begin{figure}
\includegraphics[width=9cm]{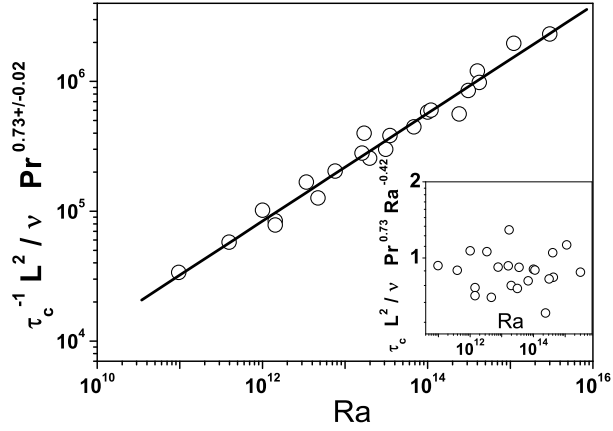}

\caption{Normalized period of oscillations of the structure functions as a function of $Pr$ and $Ra$ in both cells. The solid line is the fit $\tau_c^{-1}L^2/\nu=(0.89\pm 0.16)Ra^{0.42\pm 0.01}Pr^{-0.73\pm 0.02}$, where the coefficient is defined from the compensated plot presented in the inset.}

\label{fig:21}

\end{figure}

\begin{figure}
\includegraphics[width=9cm]{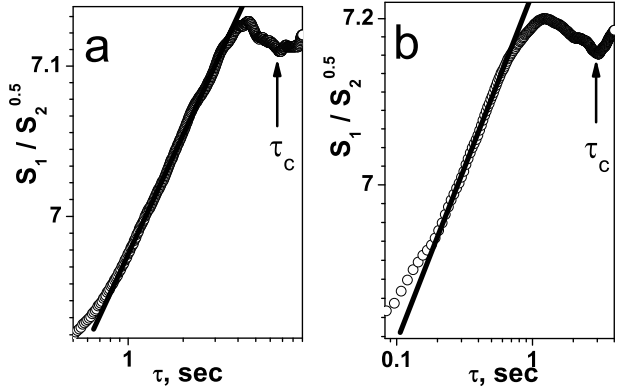}

\caption{The ratio $S_1(\tau)/[S_2(\tau)]^{1/2}$ versus time increments $\tau$ taken in the cylindrical cell at $h=45$ mm for (a) $Pr=11$ and $Ra=4\times 10^{11}$; (b) $Pr=74$ and $Ra=1\times 10^{14}$. Solid lines show the BO scaling region.}

\label{fig:22}

\end{figure}
\begin{figure}
\includegraphics[width=8cm]{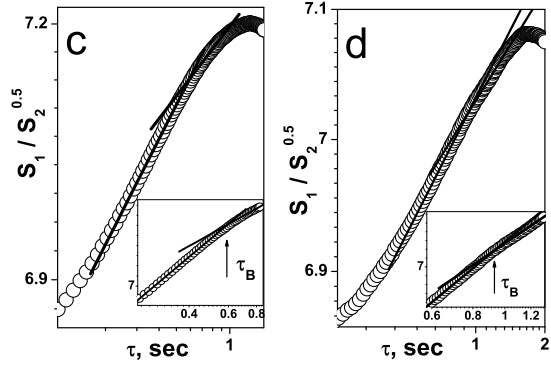}

\caption{The ratio $S_1(\tau)/[S_2(\tau)]^{1/2}$ versus time increments $\tau$ taken in the cylindrical cell at $h=45$ mm for (a) $Pr=180$ and $Ra=2\times 10^{15}$; (b) $Pr=244$ and $Ra=3\times 10^{15}$ are presented. Solid lines are the fits. Arrows in the insets show in (a) $\tau_B=0.59$ s  and (b) $\tau_B=0.93$ s. Solid lines show two different scaling regions: the BO region at $\tau>\tau_B$ and the Kolmogorov region at $\tau<\tau_B$. }

\label{fig:23}

\end{figure}

\begin{figure}
\includegraphics[width=8cm]{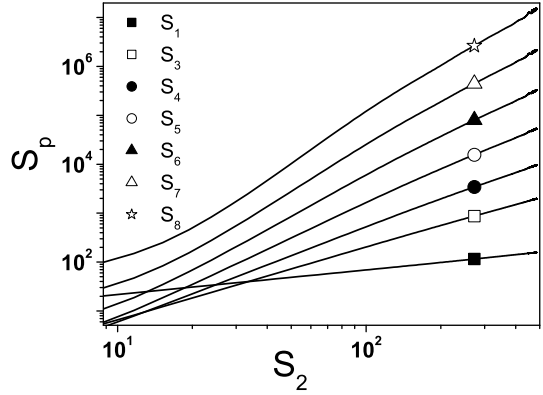}

\caption{Structure functions $S_p$ up to order $p=8$ versus $S_2$ in the ESS presentation are plotted for the data taken in the cylindrical cell at $h=45$ mm at $Pr=130$ and $Ra=2\times 10^{14}$.}

\label{fig:24}

\end{figure}

\begin{figure}
\includegraphics[width=8cm]{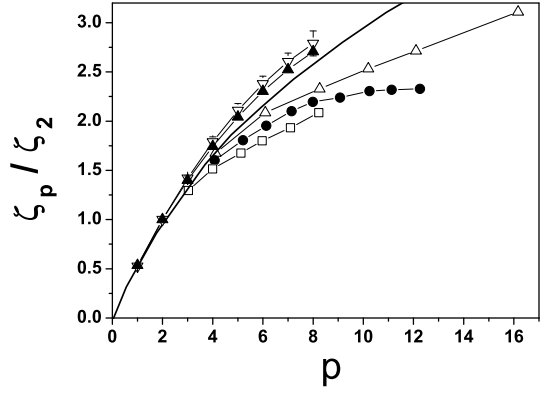}

\caption{The normalized scaling exponents $\zeta_p/\zeta_2$ versus $p$ for the data taken in the cylindrical cell at $h=45$ mm at $Pr=130$ and $Ra=2\times 10^{14}$ (solid triangles) and for the averaged over all $Pr$ and $Ra$ (open down-triangles) in comparison with the data on temperature fluctuations obtained in the experiments on turbulent convection in helium (open up-triangles) and taken from Ref. \cite{sreeni}, with the data on passive temperature taken in turbulent flows in air (open squares) from Ref. \cite{ciliberto} and in helium (solid circles) from Ref. \cite{tabeling}, and theoretical result for passive temperature fluctuations (solid line) from Ref. \cite{leveque} are shown. }

\label{fig:25}

\end{figure}

\end{document}